\documentstyle[12pt]{article}

\begin{document}

\centerline {THE NEWTON WONDER IN MECHANICS} 

\bigskip

\centerline{\it By D. Lynden-Bell}

\centerline{\it Institute of Astronomy, Cambridge}

\bigskip

\large
{Application of Newton's ideas from $Principia$ gives many {\bf new}
results in mechanics.  Here we explore the question ``What form of
extra force will maintain the {\bf magnitude} of a vector constant of
the motion while changing its direction?''}
\normalsize

\bigskip

\noindent{\it Introduction}

Shortly after his death I reviewed Chandrasekhar's last book$^{1}$
{\it Newton's Principia for the Common Reader}.  This fine book
introduced me to the whole sweep of Newton's ideas in mechanics and
sent me back to reading $Principia$ itself in
translation$^{3}$. Newton's insight stimulated many new ideas in me of
which only the following have been investigated so far.

\bigskip

\noindent{\it Revolving orbits}

In his theorum on revolving orbits Newton points out that in a
central orbit $r^2d\phi/dt = h$ then $r^2d(\alpha \phi )/dt =
\alpha h$.  So an angular motion with $\phi^\ast = \alpha \phi $
replacing $\phi $ will still have constant angular momentum.  He then
asks what extra central force is needed to leave the $r$ motion
unchanged and deduces that it must balance the extra centrifugal force
$(\alpha ^2 -1)h^2r^{-3}$.  He then deduces that the apses of such
orbits rotate at the rate $(\alpha -1)2\pi / P $ faster.  Here $P$ was
the mean period in $\phi $ of the original orbit.  The new orbit when
viewed from axes revolving at the rate $(\alpha -1)d\phi/dt $ is
exactly the same as the old one viewed from fixed axes.  But $(\alpha
-1 )d\phi/dt $ is {\bf not} a steady rotation rate.  In reference
$^{4}$ we explore the weird distorted shapes taken by such
orbits when viewed from axes rotating at the mean rate $(\alpha -1
)<d\phi/dt > $ where $< d\phi/dt > = 2 \pi/P$.

\eject

\noindent{\it Exact general N-body solutions}

Few know that Newton solved completely an N body problem for any N for
any initial conditions.  The force law for which he did this was
$${\bf F}_{{\scriptstyle IJ}} = k m_{\scriptstyle I} m_{\scriptstyle
J} ({\bf r}_{\scriptstyle J} - {\bf r}_{\scriptstyle I}).$$ The total
force on body $I$ is then directed at the barycentre at ${\bf \overline r}$
$${\bf F}_{\scriptstyle I} = \sum {\bf F}_{{\scriptstyle IJ}} =
km_{\scriptstyle I}M({\bf \overline r} -\bf r_{\scriptstyle I})$$ where $M$ is the
total mass.  Thus each body moves as though attracted to the centre of
mass which itself travels uniformly.  This force law can be found by
differentiation of the total potential energy $V = {\scriptstyle {1
\over 2}}kM^2r^2$ where

$$r^2 = \sum_{I<} \sum_J m_{\scriptstyle
I}m_{\scriptstyle J}M^{-2} ({\bf r}_{\scriptstyle J} - {\bf
r}_{\scriptstyle I})^2.$$

Ruth Lynden-Bell and I have shown$^{5}$ that this result can be
generalised to solve the non-harmonic case in which $V$ is replaced by
any $V(r)$. Some results even extend to $V$ of the form $$V_0 (r) +
r^{-2}V_2 \left ({{\bf r}_{\scriptstyle I} - {\bf r}_{\scriptstyle J}
\over r} \right )$$ where $V_0$ is an arbitrary function of $r$ and
$V_2$ is any function of its many scale-free arguments.  After some
struggles with Einstein-Bose \& Fermi-Dirac statistics we solved the
$V(r)$ problem completely in quantum mechanics$^{6}$.  We have further
work in progress on the statistical mechanics and mechanics of these
extraordinary N-body problems which can be applied to Bose-Einstein
condensates.

\bigskip

\noindent{\it Vector constants of the motion}

We all know that in the presence of a central force the radius vector
from the centre to a body sweeps out equal areas in equal times in a
plane.  However, although Newton proves it, we are not taught the
converse that if from some fixed point $S$, the radius vector to a
body sweeps out equal areas in equal times in a plane then the force
on that body is central to $S$.  This is Newton's$^{2}$ proposition
2, but he follows it with a truly remarkable scholium or rider that
no-one seems to have understood.  If we have a body subjected to a
central force but beside that there is another force which is at each
moment perpendicular to the current plane of motion, then the radius
vector to the body still sweeps out equal areas in equal times,
notwithstanding its non-coplanar motion.

To prove this consider the moment of this extra force about the
centre.  Since the extra force is parallel to the angular momentum the
moment is perpendicular to it.  Thus the rate of change of the
angular momentum is perpendicular to it, so the magnitude or `length'
of the angular momentum vector remains constant.  In the surface made
up of infinitessimal areas perpendicular to the angular momentum at
each instant the rate of sweeping of area ${\scriptstyle {1 \over 2}}
r^2d\phi / dt = {\scriptstyle {h/2}}$ is therefore constant QED.
This is all so obvious to Newton that he does not stoop to prove it -
he simply states it!

Analysing what Newton has done here, he takes a vector constant of the
motion and asks under what circumstances the magnitude of that vector
can be conserved without requiring each component to be separately
conserved.  But there are three vector conserved quantities in
classical mechanics.  They are, besides the angular momentum, the
linear momentum ${\bf p}$ and ${\bf r} - {\bf v}t = {\bf r} _0$, the
position of the barycentre at $t = 0 $.  Can we preserve $|{\bf p}|$
without keeping ${\bf p}$ constant?  Yes it requires a force $
d{\bf p}/dt=m{\bf v} \times {\bf B}$ that is always perpendicular to ${\bf p}$.
Thus we are led to forces of gravo-magnetic type of which Coriolis
force is a special example.  What forces are needed to preserve the
magnitude of ${\bf r} - {\bf v}t$?  Evidently $({\bf r} - {\bf v}t) \cdot
 {d \over dt} ({\bf r} - {\bf v} t ) = 0$ so $-{\bf \dot v}t$ must be
of the form $({\bf r} - {\bf v}t) \times {\bf B}$.

Thus the force per unit mass ${\bf F}$ must be of the form
$${\bf F} = {\bf v} \times {\bf B} - {\bf r} \times {\bf B}/t \eqno
(1)$$
where ${\bf B}$ may be any function of position, time and velocity.
Perhaps the simplest example of such a force occurs when a magnetic
field is constant in space but varies its strength with time.  The
associated electric field is ${\bf E} = {\scriptstyle {1 \over 2}}
{\bf r} \times d {\bf B} /dt$ so the total electromagnetic force on
unit charge with unit mass will be of the form (1) provided ${\bf B} =
{\bf b}t^{-2}$ with ${\bf b}$ a constant.  We orient the $z$ axis along it
so ${\bf b} = (0,\ 0,\ b)$.  Using the fact that $({\bf r} - {\bf
v}t)^2 = {\rm const} = z^2_0 + R^2_0$  the equations of motion may be
solved giving
$$
\left.
\begin{array}{rll}
x + i y = & t \left (u + w e^{ib/t}\right );\ \ x - v_xt+i(y-v_yt) =
ibwe^{ib/t}\\
z = & v_zt+\ z_0 & 
\end{array}
\right \} \eqno (2)
$$
\noindent
where $u$ and $w$ are complex constants and $v_z\  \&\  z_0$ are real
ones.  $|w| = R_0/b$.   Thus Newton's method of generalising
constants of the motion works beautifully for this problem too.

\bigskip

\noindent{\it Special relativity}

Can the idea be extended to relativity?  Here the ten classical
integrals combine into the energy-momentum 4-vector, ${\bf P}$ and the
anti-symmetric tensor 
$$K_{ij} = \epsilon_{ijkl} x^{k}{ P}^{l}\ .$$  
The invariant scalars are 
$$P^2,\ K_{ij}K^{ij}=4K^2 \ \ {\rm and} \ \
\epsilon^{ijkl}K_{ij}K_{kl}\ .$$ Unfortunately $P^2 = m^2_0c^2$ which is
always constant and the epsilon $KK$ invariant is zero identically, so
only $K^2$ is suitable for playing Newton's game.  Differentiating
with respect to comoving time $\tau $ and remembering that $m d {\bf
x}/d\tau = {\bf P}$, we find that $K^2$ is constant when the four-force 
${\bf F} = d{\bf P}/d\tau$ is perpendicular to ${\bf x} = (ct,\ x,\
y,\ z )$.  However, since $P^{2}$ is constant the four-force is always
perpendicular to ${\bf P}$.  If ${\bf N}$ is any vector along a third
independent direction perpendicular to the four-force, that force takes
the form $F_i=\epsilon_{ijkl}x^{j}P^kN^l $ so this is the general form
of force when $K^2 = ({\bf r} \times {\bf p})^2 - E^2c^{-2}({\bf r} -
{\bf v}t)^2$ is conserved.  Here $E$ is the energy.  A pretty example
is given by taking ${\bf N}$ along the time direction; then reverting
to 3-vectors and writing $N = |{\bf N}|$
$$d {\bf p}/d \tau = N {\bf r} \times {\bf p}\ . \eqno (3)$$ Under such
forces $|{\bf p}|$ is clearly constant so $|{\bf v}| = v$ is constant
too; furthermore writing ${\bf L} = {\bf r} \times {\bf p}$ we see
that $d{\bf L}/d\tau = N {\bf r} \times {\bf L}\ {\rm so}\ |{\bf L}|$ is
constant.  From our $K^2$ invariant this of course implies $|{\bf
r}-{\bf v}t|$ is constant.  ${\bf p}$ and ${\bf L}$ are perpendicular
and obey the same equation (3) which corresponds to angular rotation
at the variable rate $N{\bf r}\sqrt {1 - v^2/c^2}$.  A third
vector perpendicular to the other two is given by ${\bf R} = {\bf p}
\times {\bf L}/p^2$ which is the component of ${\bf r}$ transverse to
${\bf p}$.  It is readily shown that it too obeys $d{\bf R}/d\tau =
N{\bf r} \times {\bf R}$, so $|{\bf R}|$ is constant and ${\bf R, p,
L}$ form a triply orthogonal triad of rotating vectors of constant
length.  ${\bf v}$ is along ${\bf p}$ and is also fixed in that
rotating frame.  Now
$$d({\bf r} \cdot {\bf v})/dt = v^2 = {\rm const},$$
so
$${\bf r} = {\bf R} + {\bf v} (t + \alpha )\eqno (4)$$
since the components of ${\bf r}$ transverse to ${\bf v}$ make up
${\bf R}$.  The above equation gives the misleading impression that
the particle is travelling uniformly in a straight line.  That is
indeed true {\bf relative to the rotating axes} but in absolute axes
${\bf R}$ and ${\bf v}$ are rotating and at a time dependent rate.
Notice that ${\bf X} = {\bf r} - {\bf v}t$ is constant in the rotating
axes so ${\bf X}$ moves on a sphere in fixed axes i.e., $|{\bf X}| = {\rm
const }$.  Viewed from the rotating axes in which ${\bf R, p}\ {\rm
and}\ {\bf L}$ are fixed constant vectors the whole motion is simple
and in those axes even their rotation rate takes the simple form,
$$\mbox{\boldmath$\Omega$} = (1 - v^2/c^2)^{{\scriptstyle {1 \over 2}}}N
{\bf r} = (1 - v^2/c^2)^{{\scriptstyle {1 \over 2}}} N\left [{\bf R} +
{\bf v}(t + \alpha) \right ] \eqno (5)$$
although it varies in magnitude and direction.  Of course from the
viewpoint of a co-rotating observer the universe rotates at rate
$-\mbox{\boldmath$\Omega$}$.  The true complication of the motion is only
comprehended when one imagines what is seen by an inertial observer to
whom the inclined offset straight line, down which the motion occurs
uniformly, whirls around a changing axis at a non-uniform rate.  So far
the reader may have assumed that $N$ was a constant but nowhere have
we assumed this and indeed the force law can be such that $N$ is any
scalar function of ${\bf r, v}$ and $t$.

In parting we notice that this is but a very special case of the force
laws that preserve $K^2$.  We chose it because the others necessarily
involve $t$ explicitly but it is peculiar in that ${\bf N}$'s
direction is fixed (along the time direction) and not varying as it
would in the general case.  (3) is in fact a special case of (1) with
${\bf B}$ radial.

\bigskip
\
\noindent{\it The eccentricity vector and magnetic monopoles}

To return to where Newton started, his equation of motion for a
particle moving under a central force $-V'\hat {\bf r}$ with an extra
force $N {\bf r} \times {\bf v}$ reads for unit mass
$$d {\bf v}/dt = - V' \hat {\bf r} + N {\bf r} \times {\bf v} \eqno
(6)$$
with the central force present can we still find some rotating axes in
which the motion is especially simple?  Letting dots denote rates of
change with respect to the moving axes: $d{\bf v}/dt = \dot {\bf v} +
\Omega \times {\bf v}$.  Evidently we should again take
$\mbox{\boldmath$\Omega$} = N {\bf r}$.  Then $d{\bf r}/dt = \dot {\bf
r} = {\bf v}$ so even when $N$ varies
$$\ddot {\bf r} = - V' \hat {\bf r} \eqno (7)$$
so if we take the rotating axes the whole dynamics is the same as if
the $N$ force were absent and we had fixed axes.  Thus in the
rotating axes we have ${\bf r} \times \dot {\bf r} = {\bf h} = \ {\rm
constant \ relative \ to \ the \ rotating \ axes\ }$ and when $V = -
GM/r$ we write $\hat {\bf r} = {\bf r}/r$ and obtain
$$
\left.
\begin{array}{lllr}               
& {\bf h}\times \ddot {\bf r} = & - GM \left (\hat {\bf r} \times {\dot{\bf r}
\over r}\right )\times \hat {\bf r} = -GM {\dot {\hat {\bf r}}} & \\ 
{\rm so} & & & \\ 
&
{\bf h} \times \dot {\bf r} = & - GM(\hat {\bf r} + {\bf e})& 
\end{array}
\right \}\hfill \eqno (8)
$$
where ${\bf e}$ is a vector constant of integration.  ${\bf e}$ is
constant in the rotating axes not the fixed ones.  Those with no
history call it the Runge Lenz Vector\footnote{Hamilton$^{7}$ gave
the first coordinate free derivation in 1845 by writing the equations
in quaternion form.  Bernoulli$^{8}$, Laplace$^{9}$ and possibly
even Newton got the result earlier.  Atomic Physicists learned the
result from Runge \& Lenz much later.  See Goldstein$^{10}$} but to
me it is Hamilton's eccentricity vector pointing to pericentre with
magnitude equal to the eccentricity.  Evidently
$${\bf e} = \dot {\bf r} \times {\bf h} /(GM) - \hat {\bf r} = {\bf v}
\times {\bf h} / (GM) - \hat {\bf r} \eqno (9)$$ again ${\bf e}$ is
constant in our rotating axes.  ${\bf h}$ is ${\bf not}$ constant in
the fixed axes but obeys $d{\bf h}/dt = N{\bf r} \times {\bf h}$. 

We are now in a position to ask a question that crossed my mind but
which I would not have remembered to investigate had not S Aarseth
reminded me.  What is the most general force law that leaves $|{\bf
e}|$ invariant while letting ${\bf e}$ change?

Evidently the example given above has this property since ${\bf e}$
precesses with the axes at the rate 
$\mbox{\boldmath$\Omega$} =
N{\bf r}$ which may well be variable in magnitude as well as
direction; however that example does not give the general case.  We
now revert to fixed inertial axes.  Again ${\bf h} = {\bf r} \times
{\bf v}$ whether that quantity is constant or not and ${\bf e}$ is
defined by (9) with ${\bf v}$ written for $\dot {\bf r}$.  

Differentiating $GM{\bf e}$ we have from (9)
$$GM {d {\bf e}\over dt} = {d{\bf v}\over dt} \times {\bf h} + {\bf v}
\times \left ({\bf r} \times {d{\bf v} \over dt}\right ) - GM {d\hat
{\bf r} \over dt} \ \ . $$ The extra force per unit mass that causes
the change in ${\bf e}$ must be \linebreak${\bf F} = {d{\bf v} \over
dt} + GM r^{-2} \hat {\bf r}$,\ so
$$GM {d {\bf e} \over dt} = {\bf F} \times {\bf h} + {\bf v} \times
\left ({\bf r} \times {\bf F}\right ) \eqno (10)$$
We want this extra force to leave $e^2$ unchanged so there must be no
component of $d{\bf e} /dt$ along ${\bf e}$.  This condition may be
put in the form
$${\bf F} \cdot \mbox{\boldmath$\ell$} = 0 \eqno (11)$$
where 
$$\mbox{\boldmath$\ell$} = {\bf h} \times {\bf e} + ({\bf e} \times {\bf v})
\times {\bf r}\ \ . \eqno (12)$$ 
Thus the general ${\bf F}$ that leaves $e^2$ invariant is of the form
$${\bf F} = {\bf k} \times \mbox{\boldmath$\ell$ }\eqno (13)$$
and it will cause a change of ${\bf e}$ given by
$$GM\ d{\bf e}/dt = ({\bf k} \times \mbox{\boldmath$\ell$}) \times
{\bf h} + {\bf v} \times \left ({\bf r} \times ({\bf k} \times
\mbox{\boldmath$\ell$})\right )\eqno (14)$$ since the last term of
$\mbox{\boldmath$\ell$}$ itself involves two cross products we see
that the final term above involves five consecutive cross products and
it takes some work to sort them out and to re-express the RHS as a
cross product with ${\bf e}$ so that
$$d{\bf e} / dt = \mbox{\boldmath$\omega$} \times {\bf e} \eqno (15)$$
We write $${\bf C} = ({\bf k} \cdot {\bf r}){\bf e} + ({\bf k} \cdot
{\bf e}){\bf r}$$ and notice it is in the plane of motion and find
$$GM \mbox{\boldmath$\omega$} = 2 ({\bf k} \cdot {\bf h}){\bf h} +
({\bf v} \cdot {\bf r})\ ({\bf C} \times {\bf v})\times {\bf e}/e^2
\eqno (16)$$ Thus our result is that the general force is of the form
(13) in which ${\bf k}$ is any vector (which may vary with time) while
it leaves $e^2$ invariant. The  angular velocity of ${\bf e}$ is given by
(16).  The first term causes ${\bf e}$ to swing in the plane of motion
while the seconds swings it down through that plane (${\bf C} \times
{\bf v}$ is parallel to ${\bf h}$).  A useful final check on the
complicated derivation is that if any multiple of
$\mbox{\boldmath$\ell$}$ is added to ${\bf k}$ it must make no
difference.  Indeed ${\bf C} \times {\bf v}$ changes by
$(\mbox{\boldmath$\ell$} \cdot {\bf r}\ {\bf e} +
\mbox{\boldmath$\ell$} \cdot {\bf e \ r}) \times {\bf v}$ which is
identically zero by (12).

The general case discussed above requires complicated forces of the
form given by (12) and (13) but a great simplification occurs if we
demand that these forces do no work.  The condition ${\bf F}\cdot {\bf
v} = 0$ leads $via$ (12) \& (13) to ${\bf k}\cdot {\bf h} = 0 $ and
since we can always take ${\bf k}$ perpendicular to $\mbox{\boldmath$\ell$}$
we set ${\bf k} = N\mbox{\boldmath$\ell$} \times {\bf h} /
\mbox{\boldmath$\ell$}^2$.  Then the extra force ${\bf F}$ reverts to
Newton's simple form, $N {\bf r} \times {\bf v}$ discussed earlier
(6).  We may consider this as a magnetic force by endowing our
particle with unit charge and writing ${\bf B} = -cN{\bf r}\ \ .$ It
is then natural to require that ${\rm div}{\bf B} = 0$ so
$r^{-2}d/dr(Nr^3) = 0$ and therefore $cNr^3 = -Q = {\rm constant}$.
The resulting field is ${\bf B} = Q\hat {\bf r}/r^2$ which is clearly
the field of a magnetic monopole of strength ${\bf Q}$.  As Poincar\'e
$^{11}$ first found, motion in such fields has a singular beauty.
For a particle of mass $m$ and charge $q$ orbiting a monopole of
charge $-Zq$ and monopole strength $Q$
$$m {\ddot {\bf r}} = - \zeta \hat {\bf r} r^{-2}+ \eta
\hat {\bf r} \times {\bf v} r^{-2}$$
where $\zeta = Zq^2$ and $\eta = Qq/c$.  Cross multiplying by \ ${\bf
r},\ \ 	\ d{\bf L}/dt = - \eta {\bf h} \times \hat {\bf r}/r^2 = - \eta
(\hat {\bf r} \times {\bf v}) \times \hat {\bf r}/r = - \eta \ d \hat {\bf
r}/dt$ where ${\bf L} = m {\bf h}.$  It follows that $L^2$ is constant
and that 
$${\bf L} - \eta {\bf r} = {\bf J} = {\rm const}.$$
From which it follows that ${\bf L}^2 = {\bf J} \cdot {\bf L}$ and
$\eta = -{\bf J}\cdot \hat {\bf r}$ so ${\bf L}$ and ${\bf r}$ precess around ${\bf J}$
on coaxial cones.  The orbit is an ellipse that precesses around the
${\bf r}$ cone because of the deficit angle that is missing when the
cone is flattened, but I gave the details earlier and solved the
quantum mechanics of monopolar hydrogen.  The results here can be
extended to give the interesting motions of a small spinning sphere
both classically and in quantum mechanics $^{12}$.  Even the Dirac
equation has a pretty solution in the field of a charged monopole.  In
general relativity the equivalent of a monopole is a gravomagnetic
monopole and the required generalisation of Schwarzschild's metric is
called, NUT space, it has many weird properties.  Although the space
has the same properties when viewed from any direction its metric is
not spherically symmetric and can not be transformed into spherically
symmetric form $^{12}$.  Thus following a lead from Newton one ends
up challenging the concept of spherical symmetry in General
Relativity!  

Nouri-Zonoz and I have explored the gravitational lensing of
gravomagnetic monopoles $^{13}$ and he has found the cylindrically
symmetrical space corresponding to a line gravomagnetic monopole
$^{14}$ - the metric can not be put into a single-valued
cylindrically symmetric form!

What we have found so far comes from a tiny fraction of $Principia$; we
hope that it will encourage others to read Chandrasekhar's last book
and Newton's great one!

\eject

\centerline{\it References}

\noindent (1) S. Chandrasekhar, {\it Newton's Principia for the Common Reader}
(Oxford University Press), 1995.\\

\noindent (2) I. Newton, {\it Principia} (R.Soc., London), 1687.\\

\noindent (3) F. Cajori, {\it Newton's Principia, (Motte's Translation Revised)}
(University of California Press, Berkeley), 1934.\\

\noindent (4) D. Lynden-Bell \& R.M. Lynden-Bell, {\it R. Soc. Notes. \& Record}, 
{\bf 51}, 195.\\

\noindent (5) D. Lynden-Bell \& R.M. Lynden-Bell, {\it Proc. R. Soc. A.}, {\bf 455},
    475, 1999.\\

\noindent (6) D. Lynden-Bell \& R.M. Lynden-Bell, {\it Exact Quantum Solutions
    to Extraordinary N-body Problems}, {\it Proc. R. Soc. A.}, accepted).\\

\noindent (7) W. R. Hamilton, {\it Proc. R. Irish Acad.}, {\bf III}, Appendix p.36, 1845.\\

\noindent (8) J. I. Bernoulli, {\it Hist. Academie Royale}, p. 523, 1712.\\

\noindent (9) P. S. Laplace, {\it Trait\'{e} de m\'{e}chanique celeste (Vol 1)}
    (Paris Chez J.B.M. Duprat), 1799.\\

\noindent (10) H. Goldstein, {\it Am. J. Phys.}, {\bf 44}, 1123, 1976.\\

\noindent (11) H. Poincar\'{e}, {\it CR Acad. Sci.}, {\bf 123}, 530, 1896.\\

\noindent (12) D. Lynden-Bell,  \& M. Nouri-Zonoz, {\it Revs. Mod. Phys.}, 
{\bf 70}, 427, 1998.\\

\noindent (13) M. Nouri-Zonoz \& D. Lynden-Bell, {\it MNRAS}, {\bf 292}, 714, 1997.\\

\noindent (14) M. Nouri-Zonoz, {\it Class. Quantum Grav.}, {\bf 14}, 3123, 1997.\\

\end{document}